\documentclass{article}
\pdfoutput=1
\usepackage{spconf,amsmath,graphicx,amssymb}
\usepackage{bm}
\usepackage{cite}
\usepackage{csvsimple}
\usepackage{pgfplots}
\usepackage{tikz}
\usetikzlibrary{positioning}

\DeclareMathOperator{\Hermitian}{H}
\newcommand{\He}{{\Hermitian}}

\title{Model Order Selection in DoA Scenarios via Cross-Entropy Based Machine Learning Techniques}
\name{Andreas Barthelme, Reinhard Wiesmayr, Wolfgang Utschick}
\address{Professur f\"ur Methoden der Signalverarbeitung, Technical University Munich, 80290 Munich, Germany\\Email: \{a.barthelme, reinhard.wiesmayr, utschick\}@tum.de}

\begin{document}
%
\maketitle
\begin{abstract}
In this paper, we present a machine learning approach for estimating the number of incident wavefronts in a direction of arrival scenario. In contrast to previous works, a multilayer neural network with a cross-entropy objective is trained. Furthermore, we investigate an online training procedure that allows an adaption of the neural network to imperfections of an antenna array without explicitly calibrating the array manifold. We show via simulations that the proposed method outperforms classical model order selection schemes based on information criteria in terms of accuracy, especially for a small number of snapshots and at low signal-to-noise-ratios. Also, the online training procedure enables the neural network to adapt with only a few online training samples, if initialized by offline training on artificial data.
\end{abstract}
\section{Introduction}
\label{sec:intro}
Estimating the Directions of Arrival (DoA) of impinging electro-magnetic wavefronts with the help of an antenna array is a common task encountered in military as well as civil fields, such as aviation, autonomous driving or mobile communications. Over the last decades, many different DoA estimation approaches have been discussed (for basic introduction see, e.g., \cite{Krim1996}). However, all approaches share a common requirement, they need information about how many wavefronts are simultaneously impinging. The number of wavefronts defines the model order of the underlying wave propagation model that produces the received signal at the antenna array. Unfortunately, this model order is usually unknown a priori and has to be estimated from the received signals. Classical solutions to model order selection are based on so called information criteria \cite{Stoica2004}. Recently, a discussion of model order selection techniques from several other perspectives except for machine learning has been presented in \cite{Ding2018}.

For the most common DoA scenarios, model order selection methods based on a subspace decomposition of the sample covariance matrix of the received signals can be derived \cite{Wax1985}. They provide closed form solutions for the information criteria without the need to compute maximum likelihood estimates of the DoAs, which are computationally expensive. However, these approaches have a downside that we overcome with our proposed method. For all of the classical techniques, a high number of received signal snapshots or a high signal-to-noise-ratio (SNR) is necessary to obtain reliable estimates of the model order.

Our proposed method is based on machine learning. Machine learning techniques have been previously considered for model order estimation in DoA scenarios in \cite{CostaHirschauer1995,Costa1999}. There, model order selection up to a model order of 2 with a neural network with only one hidden layer is discussed. The authors propose a MSE criterion to train the network, where each class is mapped to a point in the two dimensional plane, similar to a modulation scheme in communications. Another approach using support vector machines has been presented in \cite{Yamamoto2003} for the closely related sound source separation problem. 

Motivated by the superiority of the cross-entropy objective for training neural networks in classification tasks \cite{Lecun1998}, we discuss using state of the art neural network architectures to solve the model order selection problem that can easily cope with an arbitrary number of sources. Additionally, we investigate how well neural networks trained on artificial data from an ideal system model can be used as an initialization for an online training approach that adapts the network to the ever present imperfections of antenna arrays in real world deployments. These imperfections are usually accounted for by calibration measurements to obtain the actual array manifold \cite{Tuncer2009}, which are still necessary, if classical DoA estimators are used. However, with a neural network as a DoA estimator, as has been previously used in the context of sound source localization \cite{Takeda2016,Li2018}, determining the actual array manifold via calibration may be replaced by online training steps similar to \cite{Takeda2016}.

\section{System Model}
\label{sec:sysmodel}
We consider a scenario where an antenna array with $M$ antennas is used to determine the number of impinging planar wavefronts that originate from $L$ sources in the far field of the antenna array. Hereby, the DoAs of the planar wavefronts are gathered in $\bm{\theta}$. In this case, the received signal $\bm{y}(t)$ at time slot $t$ reads as \cite{Krim1996}
\begin{equation}
	\bm{y}(t) = \bm{A}(\bm{\theta})\bm{s}(t)+\bm{n}(t),\label{eq:sysmodel}
\end{equation} 
where $\bm{A}(\bm{\theta})\in\mathbb{C}^{M\times L}$ is the steering matrix that depends on the array geometry and the directions of arrival $\bm{\theta}$, $\bm{s}\sim\mathcal{CN}(\mathbf{0},\sigma_s^2\mathbf{I})$ denotes the transmit signals, and $\bm{n}\sim\mathcal{CN}(\mathbf{0},\sigma_n^2\mathbf{I})$ is some additive white Gaussian noise. Throughout this work, we assume that the signals are uncorrelated with equal power $\sigma_s^2=1$. The signal-to-noise-ratio (SNR) is, therefore, given by the ratio $1/\sigma_n^2$. 

To estimate the number of wavefronts $L$, i.e., the model order, we jointly process the received signals from $N$ snapshots.

\section{Neural Network}
\label{sec:nnet}
\subsection{Data and preprocessing}
We work with artificial data created using the system model in (\ref{eq:sysmodel}), hence, an arbitrary amount of samples can be drawn for training and validation. One sample consists of $N$ received signal realizations $\bm{y}(t),\;t=1,\dots,N$. For training, we continuously sample the data, which means that in each step of the gradient descent the learning algorithm processes new, previously unseen realizations stemming from the system model in (\ref{eq:sysmodel}). As a consequence, the learning algorithm is inherently robust towards overfitting.

As the received signals are complex by nature, the samples have to be preprocessed before they can be passed to the neural network. In this work, we compare two different kinds of preprocessing. One approach is splitting up the samples in their real and imaginary parts, which are then stacked to obtain one real valued vector per sample. For the second approach, we first compute the sample covariance matrix of the received signals, i.e.,
\begin{equation}
	\bm{C}=\frac{1}{N}\sum\limits_{t=1}^N\bm{y}(t)\bm{y}^\He(t).
\end{equation}
Afterwards, we stack the real parameters of the hermitian sample covariance $\bm{C}$, i.e., its diagonal elements and the real and imaginary parts of its upper triangle, to again obtain one real valued vector per sample. Note that the input size of the neural network thus depends on the type of the employed preprocessing. Whereas the input size for the first preprocessing method is $2MN$, the covariance approach results in a input size of $M^2$ and is independent from the number of snapshots $N$.

The label for each of these samples are the respective one-hot encoded model orders of their generating system models. This means that we are considering a supervised learning approach.

\subsection{Architecture and cost function} 
We employ a fully connected feedforward neural network with $3$ hidden layers. Each hidden layer consists of $1024$ neurons, which use the rectified linear unit (ReLU) activation function. The output layer performs a softmax operation to produce $L_\text{max}+1$ outputs $z(\ell),\ell=0,\dots,L_\text{max},$ between zero and one, whose sum is again one \cite{Bridle1989}. In combination with a training based on the cross-entropy loss, these output values $z(\ell)$ can be interpreted as estimates of the posterior probabilities for each model order $\ell$ given the respective input realization $\bm{x}$. For one-hot encoded labels the cross-entropy cost function simply reduces to
\begin{equation}
	\max_{\bm{w}}\ln\left(z(\ell^*|\bm{x};\bm{w})\right).
	\label{eq:crossentropy}
\end{equation}
We can see that the training based on (\ref{eq:crossentropy}) adapts the weights $\bm{w}$ to maximize the estimate of the posterior probability of the correct model order $\ell^*$ of the input vector $\bm{x}$. In that sense, we can see the training procedure as a heuristic approach to the optimal maximum a posteriori (MAP) estimator \cite{Lecun1998}.

As the optimizer for (\ref{eq:crossentropy}), we chose the well known Adam algorithm \cite{Kingma2014} with a constant learning rate of $0.001$ and batch size of $64$.

\subsection{Initialization and online learning}
For the neural networks, we use a uniform Glorot initialization \cite{Glorot2010} to obtain a baseline for their classification performance. Additionally, we investigate how networks that have been pretrained on artificial data perform as an initialization for scenarios with slightly different system parameters and how an online learning procedure enables adaptation of the network to these changes. Such a transfer is highly relevant for an actual deployment of such machine learning approaches in a real world DoA estimator. This is due to the fact that any antenna array suffers from imperfections and its array manifold depends on the actual installation of the array due to near field scattering and mutual coupling, which has to be compensated by calibration measurements \cite[Ch.~3]{Tuncer2009}. Obtaining enough real world data to train a neural network for each installed antenna array from scratch might most often be infeasible. Therefore, we propose the aforementioned online learning procedure that is able to work on a small amount of actual measurement data due to its initialization based on artificial data. 

\section{Simulations}
In this section, we present simulation results for an uniform circular array (UCA) with $M=9$ antennas. The individual antenna elements are considered to be omnidirectional such that the array steering vector subject to the azimuth $\theta$ is given by
\begin{equation}
	\bm{a}_\text{UCA}(\theta)=\begin{bmatrix}
	\exp\{-\text{j}2\pi \frac{R}{\lambda}\cos(\theta)\}\\
	\exp\{-\text{j}2\pi \frac{R}{\lambda}\cos(\theta-\frac{2\pi}{M})\}\\
	\vdots\\
	\exp\{-\text{j}2\pi \frac{R}{\lambda}\cos(\theta-\frac{2\pi(M-1)}{M})\}\\
	\end{bmatrix},
\end{equation} 
where $R$ denotes the array radius and $\lambda$ is the wavelength of the impinging electro-magnetic wave. Note that in the formula above, we neglected the elevational dependence of $\bm{a}_\text{UCA}$, i.e., we consider all sources to lie in the same horizontal plane as the antenna array. For the following simulations, we chose the ratio of $R/\lambda$ to be $1$.

We will refer to a neural network with stacked real and imaginary part of the received signals as input data as CompNet, whereas the neural network fed with a sample covariance matrix as CovNet. To assess the performance of the proposed machine learning schemes, we compare the results to the Akaike Information Criterion (AIC) and the Maximum Description Length (MDL) method that infer the model order from the eigenvalues of the sample covariance matrix as described in \cite{Wax1985}.

If not stated otherwise, the results have been obtained with a separate training set and test set of $10^6$ samples with $N=10$ snapshots. The SNR of each sample has been drawn from a uniform distribution between $1$ and $10^3$ and the azimuth angles have been drawn from a uniform distribution between $0$ and $2\pi$. Each class, from model order $0$ to $L_\text{max}=3$, appears equally often as the other model orders in the training set and test set. 

In Table \ref{tab:accuracy}, we summarize the selection accuracy of the different model order estimation methods for the whole test set and for the samples corresponding to each individual model order. The CovNet outperforms all other methods in terms of overall selection accuracy. The CompNet on the other side, performs worse than the MDL estimator. In last place is the AIC that is prone to overfitting the model order as is also reflected in the rather poor accuracy for model orders below $3$. Note that we chose the same architecture for CovNet and CompNet for simplicity and the neural network architecture has not yet been heavily tuned, i.e., further optimization of the network architectures may improve the selection accuracy.

\begin{table}[h]
	\caption{Accuracy of neural networks vs.~information criteria}\label{tab:accuracy}
	\begin{center}
		\begin{tabular}{c|c c c c}
			Classes & CovNet & CompNet & AIC & MDL\\\hline
			\begin{filecontents*}{AccuracyLearningVsICN10.csv}
Classes,CovNet,CompNet, AIC, MDL 
All,0.9725,0.9419,0.8976,0.9514
$L=0$,0.9992,0.9980,0.9559,0.9904
$L=1$,0.9962,0.9802,0.8794,0.9598
$L=2$,0.9773,0.9126,0.7838,0.8906
$L=3$,0.9171,0.8768,0.9714,0.9646
\end{filecontents*}
\csvreader[head to column names, late after line=\\]
			{AccuracyLearningVsICN10.csv}{}{\Classes & \CovNet & \CompNet & \AIC & \MDL}
		\end{tabular}
	\end{center}
\end{table}

\begin{figure}
\begin{center}
\includegraphics{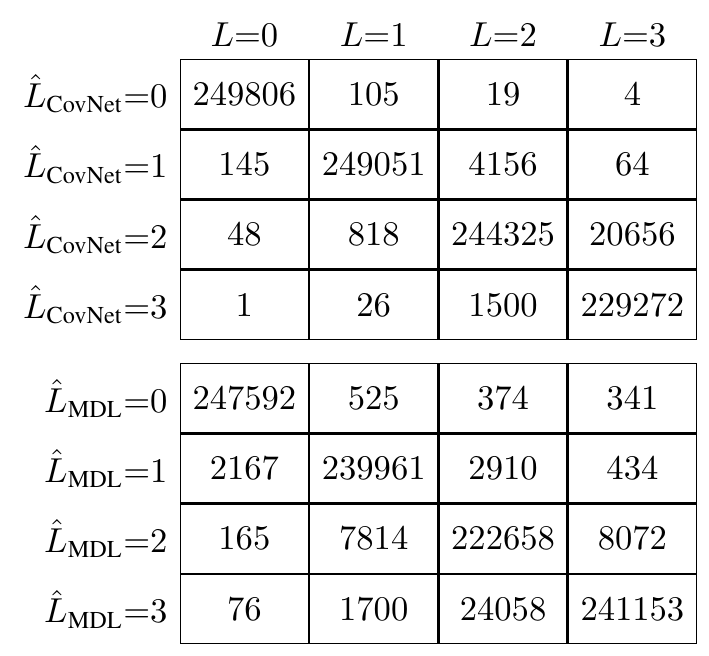}
\end{center}
\caption{Confusion matrix for CovNet and MDL.}
\label{fig:confmat}
\end{figure}

A more in depth look on the kind of model order selection errors of the CovNet and MDL estimator is presented in the confusion matrix in Fig.~\ref{fig:confmat}. There, we can see that the MDL estimator is prone to overfitting the model order, which means the predicted model order $\hat{L}_\text{MDL}$ is larger than the true model order $L$ given at the top of each column. In contrast, the CovNet tends to slightly underfit the model order.

\begin{figure}[t]
	\begin{center}
		\includegraphics{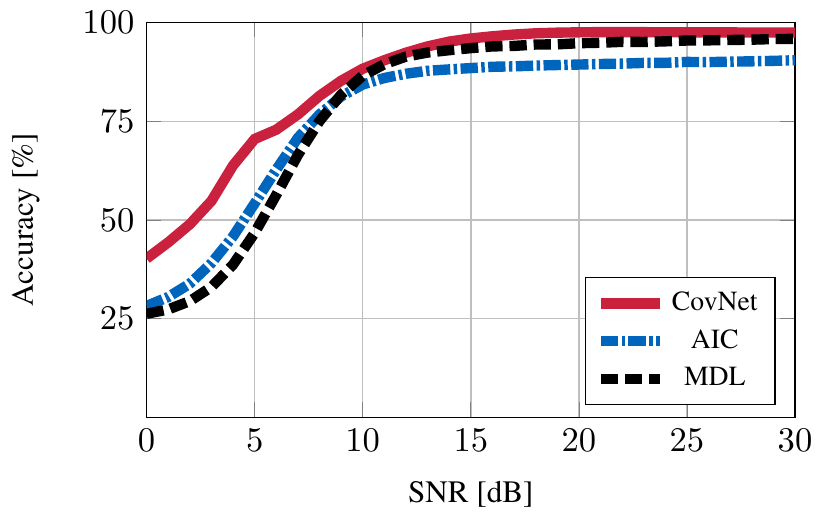}
	\end{center}
\vspace{-10pt}

	\caption{Accuracy at different SNR values.}
	\label{fig:snrcomp}
\end{figure}

To assess from which scenarios the CovNet obtains its advantage over the information criteria, we compare their performance at different SNRs in Fig.~\ref{fig:snrcomp}. We created test sets consisting of $10^5$ samples for each SNR from $0\,\text{dB}$ to $30\,\text{dB}$. The CovNet outperforms MDL and AIC over the whole SNR range. Especially at low SNR values, the neural network is able to detect the correct model order more reliably. In this region, the information criteria suffer because they inherently perform a maximum likelihood estimate of the model parameters. However, maximum likelihood estimators feature a threshold effect \cite{Athley2005}, which means that below a certain SNR threshold outliers frequently occur. Hence, the estimates are no longer unbiased as assumed in the derivation of the information criteria.

\begin{figure}[t]
	\begin{center}
		\includegraphics{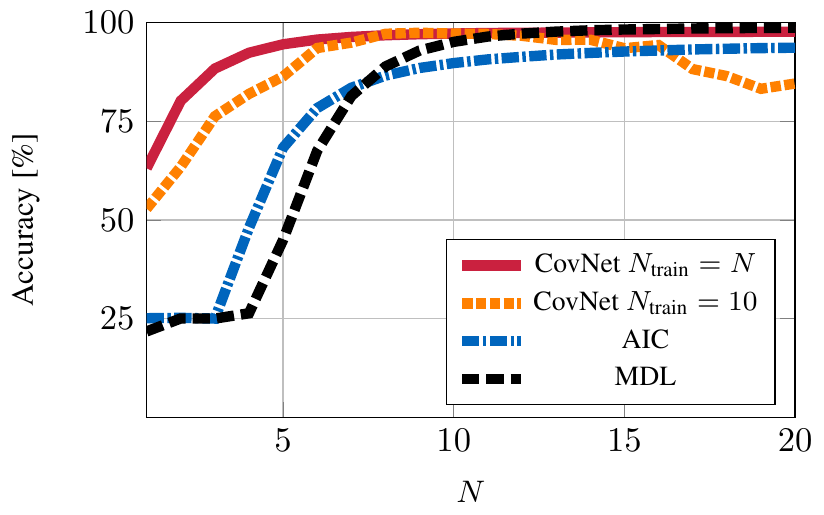}
	\end{center}
\vspace{-10pt}

	\caption{Accuracy for different number of snapshots $N$.}
	\label{fig:ncomp}
\end{figure}

Another property of interest is the accuracy for a small number of snapshots $N$. The performance in this region is, for example, very important for high-speed scanning DoA estimators or for detecting frequency hopping transmitters. Here, classical information criteria perform well, which can again be explained by the aforementioned threshold effect. In Fig.~\ref{fig:ncomp}, we compare the selection accuracy of the CovNet with MDL and AIC for a varying number of received signal snapshots per sample. We observe that the CovNet trained on the respective number of snapshots $N_\text{train}=N$ performs significantly better than the classical methods for $N<10$. For increasing $N$, the accuracy of the MDL approach improves and it even slightly beats the CovNet performance. We expect that with more effort in the architecture design of the neural network or more training data, the CovNet will be able to achieve the same performance of MDL at large $N$ values. Interestingly, the CovNet input data size is independent of the number of snapshots $N$, which allows us to evaluate the network performance
of a network trained at a certain number of snapshots $N_\text{train}$ on test sets with different $N$. As shown in Fig.~\ref{fig:ncomp}, the CovNet trained on $N_\text{train}=10$ achieves a similar accuracy as the CovNet trained on a matching number of snapshot, if $N$ does not deviate too much. Still, it is able to outperform the classical methods for small $N$, which is interesting for systems, where the number of evaluated snapshots has to be adaptable.

\begin{figure}[t]
	\begin{center}
		\includegraphics{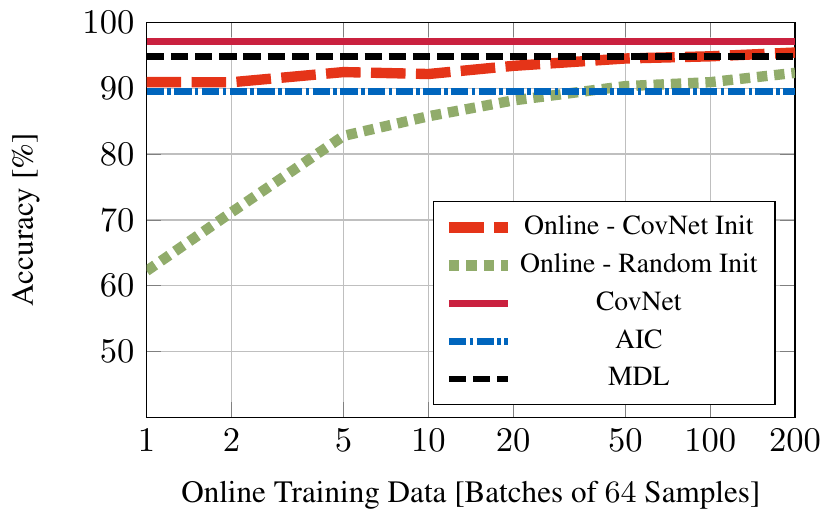}
	\end{center}
\vspace{-10pt}

	\caption{Effect of online training data, $N=10$.}
	\label{fig:onlinelearn}
\end{figure}

Finally, we investigate how well the CovNet is able to adapt to model imperfections in regard to the exact knowledge of the array manifold via online learning. We model the imperfections by the multiplication of the array manifold with a global calibration matrix $\bm{F}$ as is commonly used to model the effects of mutual coupling \cite[Ch.~3]{Tuncer2009}. The calibrated manifold then read as
\begin{equation}
	\bm{A}_\text{cal}(\bm{\theta})=\bm{F}\bm{A}_\text{UCA}(\bm{\theta}),
\end{equation}
where we exemplarily consider $\bm{F}$ to be tridiagonal with $1$ on its main diagonal and $0.25$ on the secondary diagonals, i.e.,
\begin{equation}
	\bm{F}=\begin{bmatrix}
	1&0.25&0&\hdots&0\\
	0.25&1&0.25&\hdots&0\\
	0&0.25&1&\hdots&0\\
	\vdots&\vdots&\vdots&\ddots&\vdots\\
	0&0&0&\hdots&1
	\end{bmatrix}.
\end{equation}
Fig.~\ref{fig:onlinelearn} shows the achievable accuracy of the online training approaches in relation to the available measurement data for the online training. We see that the initialization of the neural network by training on artificial data from the ideal UCA data model (CovNet Init) is superior to a random initialization. With only $1$ batch of measurement data, i.e., $64$ samples, the properly initialized network already achieves an accuracy of over $90\%$, and by that, it already outperforms AIC. $100$ or more batches of measurement data lead to an accuracy exceeding the MDL performance. There, the gap is about $1.5$ percentage points to the CovNet trained on $10^6$ samples - or $15{,}625$ batches.

\section{Conclusion}
We have shown that machine learning approaches based on the cross-entropy objective are well suited for model order selection in DoA scenarios. Especially for low SNR and a small number of snapshots, the proposed neural networks are able to surpass classical information criteria. Although each network has to be trained for a specific data generating model, we showed that neural networks with sample covariance data as their input are quite robust considering changes in the number of collected received signal snapshots. For further adaptations to imperfect knowledge of the array manifold, the presented online learning procedure showed promising results. It may be considered as a replacement for calibration measurements of the array manifold, if the subsequent DoA estimation algorithm does not rely on this information, as is the case for the machine learning based estimators.

\vfill\pagebreak

\bibliographystyle{IEEEbib}
\bibliography{MLModelOrderSelectionBib}

\end{document}